# Statistical Models for the Inference of Within-person Relations: A Random Intercept Cross-Lagged Panel Model and Its Interpretation

Satoshi Usami
Graduate School of Education, The University of Tokyo

# Statistical Models for the Inference of Within-person Relations: A Random Intercept Cross-Lagged Panel Model and Its Interpretation

The cross-lagged panel model (CLPM) has been widely used, particularly in psychology, to infer longitudinal relations among variables. At the same time, controlling for between-person heterogeneity and capturing within-person relations as processes of within-person change are regarded as key components to causal inference based on longitudinal data. Since Hamaker, Kuiper, and Grasman (2015) criticized the CLPM for its limitations in inferring within-person relations, the random intercept cross-lagged panel model (RI-CLPM), which incorporates stable trait factors representing stable individual differences, has rapidly spread, especially in psychology. At the same time, although many statistical models are available for inferring within-person relations, the distinctions among them have not been clearly delineated, and discussions over the interpretation and selection of statistical models remain active. In this paper, I position the RI-CLPM as one useful method for inferring within-person relations, explain its practical issues, and organize its mathematical and conceptual relationships with other statistical models, as well as potential problems that may arise in their application. In particular, I point out that a distinctive feature of the stable trait factors in the RI-CLPM, in representing between-person heterogeneity, is the assumption that they are uncorrelated with within-person variability, and that this point serves as an important link to the mathematical relationship with the dynamic panel model, another promising alternative.

**[Keywords] Longitudinal data, Within-person relation, Causal inference, Cross-lagged panel models, Structural equation modeling**

## Introduction

*Descriptive research and CLPM*

There are increasingly many longitudinal studies each year, with more than ten thousand papers now being published annually worldwide. Most quantitative longitudinal studies reported in developmental psychology and related fields are positioned as descriptive or exploratory research (Hamaker, Mulder, & van IJzendoorn, 2020). Descriptive studies aim to understand actual trajectories (patterns) of changes in development, growth, etc., or individual or group differences, and latent growth models, latent growth mixed models, and hierarchical linear models (also called mixed effects models, multilevel models, etc.) are often used as statistical models. On the other hand, explanatory research aims to identify the causes that bring about change and to estimate their effects quantitatively; that is, it is research oriented toward causal inference.

A core theme in causal inference is the issue of confounding, a problem that is particularly prevalent in surveys and observational research. While there is ongoing debate over how to define confounders (VanderWeele & Shpitser, 2013), here we define a confounder as "a variable that, when inferring an effect from an independent (explanatory) variable *X* to a dependent (objective) variable *Y*, affects *Y* and also affects *X* (or is covariant and has covariance with *X*)." There are two types of confounders: those whose values can change across time points (time-varying) and those whose values remain the same (time-invariant). In the statistical analysis of longitudinal studies, effects explained by past measurement data (lagged variables) and unit effects (or individual effects) as latent variables are introduced into the statistical model. By controlling for confounders and, more broadly, unobserved heterogeneity, such procedures allow closer examinations of causal relations between variables.

A common type of explanatory research in developmental psychology research is inferences of reciprocal relations, which are relations between two or more time-varying variables. For example, we might hypothesize that improved sleep habits affect mental health, or conversely that improved mental health affects sleep habits, or that both relations exist. The cross-lagged panel model (CLPM) is a statistical model that has been widely used since the 1990s to make such inferences, particularly in psychology. The CLPM is positioned as a regression model that controls for lagged variables (a cross-lagged regression model) to address the potential issue of confounding.

One reason for the increased use of the CLPM is criticisms (e.g., Rogosa, 1980) of the use of correlations (cross-lagged correlations; e.g., using the symbols described below, correlation between $X_1$ and $Y_2$ or $X_2$ and $Y_1$) or their differences to infer reciprocal relations between two variables that were measured separately in time. In addition, the subsequent rapid development of structural equation modeling (SEM) and software for its application also contributed to the widespread use of the CLPM. Indeed, a literature review of 270 articles published since 2009 in international medical journals, including psychology journals, indicated that more than 90% of statistical models used to infer reciprocal relations utilized the CLPM (Usami, Todo, & Murayama, 2019). Thus, it can be said that the CLPM has established a firmly entrenched position in explanatory research in psychology.

Meanwhile, coinciding roughly with increased use of the CLPM, the latent growth model (LGM) or latent curve model (LCM) (Meredith & Tisak, 1984, 1990) also became popular, and several statistical models combining the LCM and CLPM were proposed in the 2000s. However, relatively few users have been aware of options beyond the CLPM. In addition, these models were not proposed with the intent of directly avoiding use of the CLPM.

### *Inference of within-person relations and the RI-CLPM*

However, relatively recently, the CLPM came under substantial criticism, marking an important turning point in both methodological and applied research. The work that triggered this shift was Hamaker et al. (2015). The main issue addressed in that paper is within-person relations[1], which are the subject of this special issue paper. Within-person relations are those found in the process of intrapersonal change, such as "when a person sleeps longer, that person becomes more mentally healthy." Within-person relations are often contrasted with group-level or between-person relations, and these are not the same. Indeed, they can be exact opposites in extreme situations, such as high-intensity exercise causing heart attacks in some people (a within-person relation), despite routine exercise lowering the risk of heart attacks in most people (a between-person relation) (Curran & Bauer, 2011). In general, within-person relations are not necessarily equivalent to causality, because they do not always adequately control for possible confounder effects, but within-person relations are nonetheless considered the core of causal inference based on longitudinal data. Hamaker herself long argued for the importance of inference of within-person relations, as in Hamaker (2012) and other works.

The main criticism of Hamaker et al. (2015) is that between-variable relations that the CLPM can infer are a mixture of between- and within-person relations, and thus do not purely reflect within-person relations. As one statistical model for estimating within-person relations, they present the random-intercept CLPM (RI-CLPM), in which stable trait factor is added to the CLPM as a random-intercept that represents stable individual differences.

That paper was published in a special "longitudinal topics" issue of *Psychological Methods*, one of the leading international journals in psychometrics, and use of the RI-CLPM expanded explosively, particularly from Europe, where Hamaker and authors were based. In fact, at the time of writing this paper (August 2022), their paper had been cited more than 1,600 times, indicating the magnitude of its impact. We can expect applications of the RI-CLPM to increase even further in the future, replacing use of the CLPM.

There are only minor mathematical differences between the CLPM and the RI-CLPM, but both models are empirically known to often produce significantly different results in terms of direction ($X{\rightarrow}Y$, $Y{\rightarrow}X$, or both), effect size, and sign (positive or negative) in inferring reciprocal relations. This practical aspect also plays a role in the critique by Hamaker et al. (2015) strongly impacting psychology researchers (especially those investigating development and personality). Some attempts have been made to compare the results of data analysis and secondary analysis using the RI-CLPM and other methods (e.g., Usami, Todo, et al. 2019; Orth, Clark, Donnellan, & Robins, 2021). Then, a short time after its popularization in Europe, applications of the RI-CLPM, which is intended to infer reciprocal relations as processes of within-person change, gradually increased in psychology and related research fields in Japan.

However, as described above, there are already many statistical models that can describe reciprocal relations (as a process of within-person change) that have been proposed in psychological research and related fields, and these models are usually estimated through SEM to evaluate their goodness of fit. There are thus various ways to specify and control unobserved confounders and furthermore unobserved heterogeneity. In addition, while within-person relations and matters related to their inference have been particularly active in psychology in recent years, when we look at research fields such as economics, inferences about reciprocal and other bidirectional relations are not necessarily intended, but there are statistical models that are closely related, such as the dynamic panel data model (DPM) (Chigira, Hayakawa & Yamamoto, 2011; Wooldridge, 2011; Hsiao, 2014). Conventional differences in applied statistical models and estimation methods are thus often observed in different research areas (e.g., Hamaker & Muthén, 2020). That being the case, is application of the RI-CLPM always justified? If so, why? As a related question, from the perspective of causal inference, what is the significance of controlling for stable trait factors as stable individual differences? These questions were not necessarily sufficiently addressed in Hamaker et al. (2015).

While the author himself has, to some extent, investigated the comparison and selection of statistical models used to infer reciprocal relations (Usami, Murayama, & Hamaker, 2019; Usami, 2021), in part due to the many types of statistical models, many matters related to these points are still being debated, and psychometrics researchers do not always share a common view. In fact, criticisms of the RI-CLPM have recently been raised (Lüdtke & Robitzsch, 2022), and alongside the diversity of assumed data generating processes and actual research hypotheses, the debate over methodologies for inferring reciprocal relations as within-person relations can be considered to have become more diverse and more complex. To the author's knowledge, there has been no literature review of such topics, including an overview of the RI-CLPM.

*Structure of this paper*

The remainder of this paper is structured as follows. We first describe the model representation of the CLPM and the RI-CLPM and some analysis examples. Next, from the standpoint of positioning the RI-CLPM as an effective method for estimating reciprocal relations as within-person relations, we discuss and explain some practical topics related to applying the RI-CLPM (numbers of required time points, index for variance of stable trait factors, model extensions, measurement error assumptions, and improper solutions). Then, based on recent discussions, including the author's own studies (Usami, Murayama et al., 2019; Usami, 2021, 2023), we summarize problems that can arise when inferring within-person relations, along with an overview of other statistical models. We next discuss the DPM as another major alternative to the RI-CLPM, together with considerations regarding model selection. In particular, we point out that a distinctive feature of the RI-CLPM, in terms of representing between-person heterogeneity, is its assumption that the trait factor is uncorrelated with the within-person relations, and that this assumption also constitutes an important point of contact linking the RI-CLPM to the DPM. Finally, we summarize this paper and discuss prospects for future research.

## CLPM and RI-CLPM representations and analysis examples

*Representation of the CLPM*

In the following, we describe each model with assuming typical situations where the CLPM or RI-CLPM are used. Our model notation follows Usami, Murayama, et al. (2019).

Suppose that two continuous variables $X$ and $Y$ are simultaneously and longitudinally measured several times, and that we are interested in inferring a reciprocal relation between those variables. As described below, the number of time points $T$ is typically around 2 to 6, and only rarely 10 or more[2]. For measurements $x_{it}$ and $y_{it}$ at discrete time point $t$ $(1 \cdots t \cdots T)$ for individual $i$ $(1 \cdots i \cdots N)$, the CLPM is expressed as

$$x_{it} = \alpha_{xt} + \beta_{xt} x_{i(t-1)} + \gamma_{xt} y_{i(t-1)} + d_{xit},$$
$$y_{it} = \alpha_{yt} + \beta_{yt} y_{i(t-1)} + \gamma_{yt} x_{i(t-1)} + d_{yit}, \qquad (1)$$

for $t \geq 2$. Here, $\alpha_{xt}$ and $\alpha_{yt}$ are cross-lagged regression intercept terms, and $\beta_{xt}$ and $\beta_{yt}$ are autoregressive coefficients that express the degree to which the variables (ordered as $x_{it}, y_{it}$) in the present ($t$) can be explained from the same lagged variables (ordered as $x_{i(t-1)}, y_{i(t-1)}$), which are measurements from the past ($t - 1$). Autoregressive coefficients in this equation are said to be of the first order because we are considering regression to the lagged variable of the previous time point. Similarly, $\gamma_{xt}$ and $\gamma_{yt}$ are first-order cross-lagged coefficients that represent the degree to which the lagged variable (ordered as $y_{i(t-1)}, x_{i(t-1)}$) can explain another variable in the present (ordered as $x_{it}, y_{it}$). Cross-lagged coefficients are core parameters for estimating reciprocal relations: the larger their absolute value, the stronger the relation between the corresponding variables.

By controlling for the autoregressive term, which is past information, the CLPM allows for interpretations that go beyond simple cross-lagged correlations measured at different time points. For example, $\gamma_{xt}$ would be a quantity representing the increment of the current $x_{it}$ expected when a past $X$ measurement $x_{i(t-1)}$ is controlled (i.e., assuming a hypothetical population whose values are expected to be the same) and the past $Y$ measurement $y_{i(t-1)}$ increases by one unit ($y_{i(t-1)} \Rightarrow y_{i(t-1)} + 1$). $d_{xit}$ and $d_{yit}$ are residual terms of the cross-lagged regression, which are usually assumed to follow a bivariate normal distribution (with a zero mean vector), and the time-varying residual covariance is estimated. The measurements ($x_{i1}, y_{i1}$) at the first time point ($t = 1$) are treated as exogenous variables that do not result from the other variables, and their mean and (co)variance are assumed and estimated.

In Eq. (1), the subscript $t$ reflects the assumption of time-varying quantities for intercepts, autoregressive coefficients, cross-lagged coefficients, and other parameters. This reflects the fact that in typical longitudinal studies applying the CLPM, it is not uncommon for intervals between measurement time points to be on the scale of months or years, which may lead to qualitative differences in the dynamics of change between measurements, and further that those intervals are not always equally spaced. Of course, assuming whether time-varying or time-invariant parameters will depend on the study and be determined by the analyst according to factors such as the nature of the variables being examined, the measurement period, the type of research hypothesis (such as whether the main interest is to examine temporal stability in dynamics of changes), and the estimation-related reasons (such as a desire to reduce the number of free parameters to make the estimation more stable). While the information criterion and other general-purpose methods can be applied to model selection, since the CLPM is (assuming a multivariate normal distribution) usually estimated through SEM, various model fit indices can also be referred to (see, e.g., Toyota (1998) and Kline (2016) for the basics). Furthermore, it is highly worthwhile to examine the magnitudes of each element of the residual correlation matrix (Kline, 2016).

As an alternative to Eq. (1), the CLPM can be expressed using the deviation from the group mean at each time point, as follows (e.g., Hamaker et al., 2015):

$$x_{it} = \mu_{xt} + x_{it}^*, \quad y_{it} = \mu_{yt} + y_{it}^*, \quad (2a)$$
$$x_{it}^* = \beta_{xt} x_{i(t-1)}^* + \gamma_{xt} y_{i(t-1)}^* + d_{xit}^*,$$
$$y_{it}^* = \beta_{yt} y_{i(t-1)}^* + \gamma_{yt} x_{i(t-1)}^* + d_{yit}^*, \quad (2b)$$

where $\mu_{xt}$ and $\mu_{yt}$ are the group means at time $t$, and $x_{it}^*$ and $y_{it}^*$ are the deviations of individual $i$ from those means. The mean of the deviation is 0. In Eq. (2b), autoregressive terms, cross-lagged terms, and residuals are specified for the deviations, as in Eq. (1); however, reflecting the fact that the deviations have a mean of zero, the intercept term α included in Eq. (1) is not included. In other words, the difference between Eqs. (1) and (2) can be described as the difference between expressing the mean structure of each variable by the intercept term α or the group mean μ, where μ can be expressed as a function including $\alpha$ (Usami, 2021). By contrast, autoregressive coefficients, cross-lagged coefficients, and residual (co)variance are the same, regardless of such differences in expression. Figure 1a shows a path diagram for the CLPM when using the representation based on Eq. (2).

While the mathematical differences between these representations are trivial, as discussed below they are useful in distinguishing between existing statistical models in a conceptual, mathematical sense (e.g., Usami, Murayama et al., 2019).

*Representation of the RI-CLPM*

The cross-lagged coefficient γ in the CLPM was a quantity reflecting reciprocal relations between variables after controlling for the autoregressive term (e.g., for $\gamma_{xt}$, assuming a hypothetical population with same $x_{i(t-1)}$ values). That being the case, can we consider these cross-lagged coefficients as quantities reflecting reciprocal relations as within-person relations? It can be said that the inclusion of autoregressive terms allows the model to represent relations that go beyond the group-level relations; however, these should be understood merely as correlations within a subpopulation defined by having the same value at the previous time point, and in that sense they cannot be regarded as quantities representing within-person relations. In other words, in most cases, it is unlikely that unobserved heterogeneity is adequately controlled.

Hamaker et al. (2015) criticized the CLPM as an inappropriate method for inferring within-person relations, arguing that although it is necessary to control for stable trait factors representing stable individual differences in order to draw inferences about within-person relations, the CLPM does not take such components into account, and thus proposed the RI-CLPM as an alternative statistical model.

In the RI-CLPM, the measurements $x_{it}, y_{it}$ are first split in a manner similar to Eq.

(2a), as

$$x_{it} = \mu_{xt} + I_{xi} + x^*_{it}, \qquad y_{it} = \mu_{yt} + I_{yi} + y^*_{it}. \qquad (3a)$$

As in the CLPM, $\mu_{xt}$ and $\mu_{yt}$ are the group means at time $t$. Also, $I_{xi}$ and $I_{yi}$ are stable trait factors representing stable individual differences for individual $i$, and their (co)variances are assumed and estimated. This factor covariance reflects a between-person relation that is stable over time.

Assume that $x^*_{it}$ and $y^*_{it}$ are deviations for an individual $i$ and are uncorrelated with the stable trait factors. The group means of the stable trait factor and the deviation are both 0, and the expected scores for $x_{it}$ and $y_{it}$ (conditioned on the stable trait factor) become $\mu_{xt} + I_{xi}$ and $\mu_{yt} + I_{yi}$. By controlling for the stable trait factor as a time-invariant stable individual difference, $x^*_{it}$ and $y^*_{it}$ in the RI-CLPM can be interpreted not as deviations from the group mean like in the CLPM, but as deviations from the expected scores for each individual, or as quantities representing within-person variability that are uncorrelated with stable individual differences. The latter interpretation implies that Eq. (3a) orthogonally decomposes the variance of the observed measurements into between-person variance, represented by the stable trait factor, and within-person variance, represented by within-person variability. $x^*_{it}$ and $y^*_{it}$ can then be represented as

$$x^*_{it} = \beta_{xt} x^*_{i(t-1)} + \gamma_{xt} y^*_{i(t-1)} + d^*_{xit},$$
$$y^*_{it} = \beta_{yt} y^*_{i(t-1)} + \gamma_{yt} x^*_{i(t-1)} + d^*_{yit}. \qquad (3b)$$

This is formally similar to Eq. (2b), but unlike the CLPM it represents a regression model for within-person variability. Therefore, for example, the cross-lagged coefficient $\gamma_{xt}$ can be interpreted as the expected increase in the current $x^*_{it}$ associated with a one-unit increase in that individual's $y^*_{i(t-1)}$, conditional on the individual having a previous within-person deviation $x^*_{i(t-1)}$ and a given stable trait factor score $I_{xi}$.

Figure 1b shows a path diagram for the RI-CLPM. Since the only difference between the RI-CLPM and the CLPM is the RI-CLPM's inclusion of the stable trait factor, if the (co)variance of the stable trait factor is 0 (i.e., if $I_{xi} = I_{yi} = 0$), they are mathematically equivalent. In the RI-CLPM, as in the CLPM, the within-person variability at the first time point ($x^*_{i1}$, $y^*_{i1}$) is treated as an exogenous variable, and the mean and (co)variance are assumed and estimated. Note that as in the case of the CLPM, the choice of whether to assume time-varying or time-invariant quantities for parameters such as group means, autoregressive coefficients, and cross-lagged coefficients (and whether to assume equivalence only at certain time points) is a matter for analysts to decide from the nature of the variables being examined, the measurement period, and the type of research hypothesis. When model selection is required, one can use procedures such as model fit indices, information criteria, and residual correlations.

*Comparison of estimation results and analysis examples*

If the data generating process corresponds to the RI-CLPM and the stable trait factor has nonzero variance, then using the CLPM to analyze the data might produce dramatically different results for the cross-lagged coefficients and other inferential results from what would result from use of the RI-CLPM. In particular, the larger the (co)variance of the stable trait factors, the more likely it is that the statistical significance, magnitude, and sign of the cross-lagged coefficient estimates will differ across models, and the RI-CLPM will be judged to be a better fit. There are many reports of empirical estimation results, including Hamaker et al. (2015), Usami, Murayama et al. (2019), Usami, Todo et al. (2019), and Orth et al. (2021). More specifically, those studies showed that autoregressive coefficients estimated by the RI-CLPM tend to be smaller and that standard errors for autoregressive coefficients and cross-lagged coefficients tend to be larger, reflecting the fact that the RI-CLPM considers within-person relations after controlling for stable trait factors (Mulder & Hamaker, 2020).

Generally speaking, different estimation results can obviously be obtained by applying mathematically different statistical models. However, stable trait factors that reflect stable individual differences (e.g., $I_{xi}$) contribute to each element of the variance–covariance structure of the corresponding variable ($X$) by an equal magnitude (specifically, the variance $V(I_{xi})$ of the stable trait factor) reflecting the large impact on the estimation results of the full set of parameters.

As preparation for the discussion of model selection presented later, we provide a brief empirical example aimed at comparing estimation results. Table 1 shows the results of applying the CLPM, the RI-CLPM, and other statistical models described below to longitudinal data ($T = 6, N = 4,671$) collected from the Minnesota Adolescent Community Cohort (MACC) on adolescents' perceived degree of exposure to smoking through movies ($X$) and actual smoking intensity ($Y$). We used maximum likelihood estimation based on SEM (ML-SEM) with the "lavaan" package in R (Rosseel, 2012) for the parameter estimation. The analysis code can be obtained from the author's website (http://usami-lab.com/Usami_2022_jsdp_code.docx). Note that Usami, Murayama et al. (2019) also compared the results of various statistical models using the same data, but due to their relation to the following discussion, we focused only on certain models, such as the CLPM and the RI-CLPM. Here we present the results of estimated autoregressive coefficients, cross-lagged coefficients, and residual (co)variance under time-varying or time-invariant assumptions. However, since improper solutions occurred under the time-varying conditions of the RI-CLPM, in this condition only the residual (co)variances are treated as time-invariant for the sake of comparative convenience. We explain how to deal with improper solutions below. See Choi, Forster, Erickson, Lazovich, and Southwell (2012) and Usami, Murayama et al. (2019) for further details related to the data.

The estimation results indicate that, regardless of whether time-varying or time-invariant assumptions are imposed, the estimates, statistical significance, and signs of the cross-lagged coefficients differ between the CLPM and the RI-CLPM. Estimates of autoregressive coefficients were smaller in the RI-CLPM, and standard errors for the estimates of these coefficients were larger in the RI-CLPM. In terms of model fit, the RI-CLPM showed better fit. The results obtained from applying other statistical models will be discussed later.

**Some practical issues concerning the application of the RI-CLPM**

*The number of time points required*

Both the CLPM and the RI-CLPM are usually estimated within an SEM framework, meaning the objective function is set and the solution is obtained based on making the mean structure and variance–covariance structure of the model close to the sample mean (vector) and the sample variance–covariance matrix of the observed data. Bollen (1989) and Toyoda (1998) explain the fundamentals of SEM optimization, and Toyoda (1992, 2012) provides details about estimation issues.

As a necessary condition for model identification, in other words, to obtain a unique solution, the CLPM requires longitudinal data with at least $T = 2$. By contrast, the RI-CLPM assumes a stable trait factor and estimates its variance and covariance, resulting in three more parameters. Reflecting this, the RI-CLPM requires longitudinal data with at least $T = 3$. These differences arise regardless of whether one assumes time-varying or time-invariant autoregressive and cross-lagged coefficients.

According to Usami, Todo, et al. (2019), who reviewed 270 studies in international journals in medicine, psychology, and related fields that reported inferential results about reciprocal relations, 106 studies (39%) used longitudinal data with $T = 2$, 89 (33%) used $T = 3$, 36 (13%) used $T = 4$, and 16 (6%) used $T = 5$. Only 24 (9%) used longitudinal data with $T \geq 6$. In Hamaker et al. (2015), which conducted a similar review of articles published in psychology in 2012, it was reported that nearly half (45%) of the 115 studies applying the CLPM used longitudinal data with $T = 2$. This is consistent with the results of Usami, Todo et al. (2019).

As noted above, in response to the criticism by Hamaker et al. (2015), there has been growing momentum toward comparative examinations of analytic results and secondary analyses that take the RI-CLPM into consideration; however, the RI-CLPM cannot be identified from longitudinal data with $T = 2$. Therefore, to make inference about within-person relations based on the RI-CLPM possible, it is first necessary to adopt a research design that presupposes the collection of longitudinal data with $T \geq 3$. Moreover, other things being equal, increasing the number of time points is expected to yield more stable estimates of the model parameters, especially the (co)variances of the stable trait factors. As a general matter, however, the magnitudes of autoregressive and cross-lagged coefficients will depend on the lengths of intervals between measurement points (lag; see, e.g., Dormann & Griffin, 2015), so the lag and number of time points should be determined while considering actual research hypotheses and measurement periods.

*Index for variance of stable trait factors*

As described above, in Eq. (3a) the stable trait factor and deviation (within-person variability) are not correlated, so the variance of observation at each time point (e.g., variance $V(y_{it})$ for variable *Y*) can be orthogonally decomposed as the sum of the variance of the stable trait factor (between-person variance: $V(I_{yi})$) and the variance of the within-person variability (within-person variance: $V(y_{it}^*)$).

When inferring within-person relations, the magnitude[3], statistical significance, signs, and standard errors (confidence intervals) for estimated cross-lagged coefficients, as well as coefficients of determination in lagged regressions, are of primary interest. However, the proportion of the variance of the stable trait factor to the variance of observation at time *t*:

$$R_{yt}^2 = \frac{V(I_{yi})}{V(y_{it})} = \frac{V(I_{yi})}{V(I_{yi}) + V(y_{it}^*)} \quad (4)$$

can also be a useful index. The larger the value of $R_{yt}^2$, the greater the proportion of between-person variance at each time point, and when the trajectories of multiple individuals are considered, the extent to which those trajectories are separated from one another becomes increasingly pronounced.
By contrast, the smaller the value of $R_{yt}^2$, the larger the proportion of within-person variance, so the extent to which each individual trajectory comes closer together and overlaps becomes more pronounced.

The magnitude of $R_{yt}^2$ is considered to depend on characteristics such as the nature of the variables being examined, the measurement period, and even the measurement method (e.g., reliability of the measurement). Although there is, of course, variation in degree, in many cases it is natural to assume that, in actual longitudinal data, some individuals consistently show relatively high observations, whereas others consistently show relatively low values. This suggests that the variance of the stable trait factor as a stable individual difference is nonzero. Indeed, studies applying the RI-CLPM often show variance estimates of a non-negligible magnitude, as illustrated in Table 1, and in such cases the estimated cross-lagged coefficients and related results are more likely to differ substantially between the CLPM and the RI-CLPM.

Using longitudinal data ($T = 3$; ages 10, 12, and 14) on depressive symptoms (Short Mood and Feelings Questionnaire scores), sleep durations, bedtimes, and body mass index (BMI) measured in the Tokyo Teen Cohort Study conducted to investigate adolescent mental and physical development (for the protocol paper, see Ando et al., 2019), Usami (2023) estimated proportions of variance of stable trait factors to the variance of each variable at the first time point ($R_1^2$; $t = 1$ and age 10). As a result, the estimated values obtained were, in order, $R_1^2 = 0.245, 0.545, 0.482, 0.748$. In other words, compared with the other variables, depression exhibited relatively large within-person variability (within-person variance), whereas sleep duration and bedtime as lifestyle-related variables, and particularly BMI as an indicator of body composition, appeared to be more strongly influenced by stable individual differences, with $R_1^2$ tending to be relatively large.

### *Some extensions to the RI-CLPM*

When the effects of variables are thought to persist over a longer period, one may consider specifying not only first-order but also second- or higher-order autoregressive and cross-lagged terms (e.g., effects of $x^*_{i(t-2)}$ on $x^*_{it}$ and $y^*_{it}$). If such effects are indeed present, incorporating them may also help address confounding, for example by allowing the causal effects of first-order lagged variables to be estimated more accurately, and may improve model fit.

One procedure to represent a model that includes a time-varying confounder ($z_{it}$) is to assume linear relations among the variables and to extend Eqs. (3a) and (3b) in a straightforward manner to the three-variable case involving $X$, $Y$, and $Z$. When the variable $Z$ is introduced as a mediator in such a model, the mediation effect can be estimated if various model assumptions hold (e.g., assumption of linearity among variables and appropriate control of the confounders), for example, through the series $x_1^* \Rightarrow z_2^* \Rightarrow y_3^*$ (based on the within-person relations evaluated by controlling for the stable trait factor of each variable, rather than the group-level relation that is often examined in practice).

As yet another extension of the RI-CLPM, Mulder and Hamaker (2020) illustrate and explain model representations for (a) the case where there are time-invariant observed variables, (b) the case of extension to a multi-group model, and (c) the case of dealing with multiple indicators (for example, when a construct such as depression is measured longitudinally using psychological scale one may specify a measurement model for each individual item [indicator] rather than relying on a sum score).

*Assumption of correlation between stable trait factors and within-person variability*

The RI-CLPM assumes that there is no correlation between the stable trait factor as a stable between-person difference and within-person variability. This is because there is no reason to assume that within-person variability, which is positioned as a temporal deviation from each individual's expected score (which is a function of the stable trait factor), is correlated with the stable trait factor (Usami, Murayama et al., 2019). Nevertheless, as Figure. 1c shows, it is procedurally possible to assume a correlation, and specifically for data with $T = 4$ or more, such model can be identified regardless of whether time-invariant coefficients and residual (co)variances are assumed.

Under this setting, however, the extent to which the stable trait factor contributes to the variance of the observations varies across time points. Consequently, the conceptual meaning of the stable trait factor as representing stable individual differences also changes. As a related matter, the fact that the variance of the observation is not orthogonally partitioned into the variance of the stable trait factor and the variance of the within-person variability makes it less meaningful to report the index $R^2_{yt}$ defined above. The emergence of such difficulties also suggests that, in light of the model's original intent, a specification that assumes a correlation between the stable trait factor and within-person variability can no longer, strictly speaking, be regarded as an RI-CLPM. Even so, such a specification constitutes an important point of contact that establishes a mathematical connection with the dynamic panel model (DPM; Figure 1d) discussed later (Andersen, 2022).

*Assumptions of measurement errors and improper solutions*

Among the statistical models proposed prior to Hamaker et al. (2015), one that is mathematically very close to the RI-CLPM is the stable trait, autoregressive trait, and state model (STARTS; Kenny & Zautra, 1995, 2001), which was proposed in personality research[4].

In the STARTS model, the observations $(x_{it}, y_{it})$ are partitioned into true scores $(f_{xit}, f_{yit})$ and measurement errors $(\varepsilon_{xit}, \varepsilon_{yit})$ that are assumed to be uncorrelated with those true scores as

$$x_{it} = f_{xit} + \varepsilon_{xit}, \qquad y_{it} = f_{yit} + \varepsilon_{yit}. \qquad (5a)$$

Because the measurement errors have mean zero, the mean of the observations is equal to that of the true scores. The measurement errors at each time point are usually assumed to follow a bivariate normal distribution, and the measurement error (co)variances are estimated. However, due to identification constraints, time-invariant measurement error (co)variances are usually assumed.

Then, as in the RI-CLPM, we divide these true scores at each time point into the group mean, stable trait factor, and deviation (within-person variability), and specify a regression equation for the deviation as

$$f_{xit} = \mu_{xt} + I_{xi} + f^*_{xit}, \qquad f_{yit} = \mu_{yt} + I_{yi} + f^*_{yit}, \qquad (5b)$$
$$f^*_{xit} = \beta_{xt} f^*_{xi(t-1)} + \gamma_{xt} f^*_{yi(t-1)} + d^*_{xit},$$
$$f^*_{yit} = \beta_{yt} f^*_{yi(t-1)} + \gamma_{yt} f^*_{xi(t-1)} + d^*_{yit}. \qquad (5c)$$

Although there are differences according to whether the equations are for observed or true values, the specifications for Eqs. (5b) and (5c) are basically the same as those for the RI-CLPM (Eqs. (3a) and (3b)).

Note that the STARTS model often imposes nonlinear constraints on the parameters so that, for example, the variance of each component of the stable trait factor, the within-person variability (corresponding to the "autoregressive trait" in the STARTS model), and the measurement error (similarly corresponding to the "state") satisfy stationarity[5] in the sense that they are time-invariant (Donnellan, Kenny, Trzesniewski, Lucas, & Conger, 2012). Such constraints are not usually imposed in the CLPM or RI-CLPM.

In (developmental) psychology research, researchers frequently attempt to measure latent constructs that cannot be directly observed; however, because measurement reliability is often imperfect, primarily due to the measurement methods themselves, it is not uncommon for the resulting data to contain measurement error. This would introduce bias in the parameter estimates of the RI-CLPM, which does not directly account for measurement error (especially at the initial time point).

Although the two models differ in their conceptual interpretations, it is somewhat ironic that, whereas the STARTS model was proposed before the RI-CLPM and provides a more general formulation, it is the RI-CLPM that has rapidly gained popularity in recent years. However, it is empirically known that including measurement error as in the STARTS model comes at the cost of a tendency to produce improper solutions, such as negative estimates for the variances of the stable trait factors or for the residual and error variances, as well as the failure of the variance–covariance matrix of the stable trait factors to be positive definite (e.g., Hamaker et al., 2015; Usami, Murayama et al., 2019; Usami, Todo et al., 2019; Orth et al., 2021). To address this issue, several approaches have been suggested, including increasing the number of time points and using multiple indicators (Cole, Martin, & Steiger, 2005; Luhmann, Schimmack, & Eid, 2011), as well as conducting Bayesian estimation with prior distributions specified for the parameter values (Lüdtke, Robitzsch, & Wagner, 2018). It is also empirically known that improper solutions may arise in the RI-CLPM, as in the analysis example above; however, they occur more frequently in the STARTS model (e.g., Usami, Todo et al., 2019).

In addition, from the standpoint of model identification, the STARTS model is subject to somewhat stronger constraints than the RI-CLPM. Because the STARTS model is a statistical model with a larger number of parameters and a more complex structure, longitudinal data with at least $T = 4$ waves are required for identification. To effectively avoid improper solutions and to obtain more stable parameter estimates, data with around $T = 10$ waves may be required (Kenny & Zautra, 2001). As noted above, however, studies that actually collect longitudinal data on such a scale to examine reciprocal relations are limited (Usami, Todo et al., 2019).

### *Other statistical models and their relation to the RI-CLPM*

Many other statistical models, including the STARTS model, are available to examine reciprocal relations at the within-person level, and some have been proposed after Hamaker et al. (2015). While these models were proposed in psychology and related fields and estimated usually through SEM framework, there are still other options if we look beyond psychology to other fields such as economics. This fact means that there are various procedures to model and to control for unobserved confounders, as well as unobserved heterogeneity.

While the RI-CLPM seems to provide more reasonable model representations to infer within-person relations, at least when compared with the CLPM, is application of the RI-CLPM always justified, despite the availability of so many other statistical models? If so, why? As a related issue, what is the significance of controlling for stable trait factors as stable individual differences in terms of causal inference? These issues do not appear to have been discussed in sufficient detail in Hamaker et al. (2015).

In my own work (Usami, Murayama et al., 2019; Usami, 2021), I presented a unified framework for comparing several statistical models and organizing their conceptual and mathematical relationships, and on that basis examined differences in the interpretation of cross-lagged coefficients, positioning the RI-CLPM as one useful approach for inferring within-person relations. However, due in part to the many types of available statistical models and the complex mathematical relations among them, discussion remains ongoing regarding model selection and estimation methodology especially from the perspective of causal inference. The issue of inferring reciprocal relations as within-person relations is thus becoming increasingly complex and diverse. Even when the discussion is limited to comparisons between the CLPM and the RI-CLPM, there are currently contrasting views: some argue for the usefulness of the CLPM (that includes second- or higher-order lags) (Lüdtke & Robitzsch, 2022), whereas others, like Hamaker et al. (2015), have once again provided a critical reappraisal of the CLPM (Lucas, 2023).

Another important factor that complicates the discussion is the difficulty of determining whether common factors included in statistical models, such as stable trait factors, adequately control for unobserved confounders (and furthermore, unobserved heterogeneity), despite the actual data generating process and the true model being generally unknown to the researcher. This makes it difficult to draw a clear-cut conclusion regarding the best selection of model for inferring within-person relations and how cross-lagged coefficients should be interpreted.

In this section, from the standpoint that regards the RI-CLPM as one useful approach for inferring within-person relations, and drawing on recent discussions including the author's own work (Usami, Murayama et al., 2019; Usami, 2021, 2023), I discuss three issues: (1) an overview of several statistical models within the SEM framework that have been proposed in psychology and related fields (i.e., the LCM-SR, LCS, and GCLM), together with the inferential problems that may arise when using them; (2) an overview of the dynamic panel model (DPM), used primarily in economics, and its mathematical relationship to the RI-CLPM; and (3) the role of stable trait factors in the RI-CLPM from the view of causal inference, as well as the issue of model selection. In particular, I point out that a distinctive feature of the stable trait factor in the RI-CLPM, in terms of representing between-person heterogeneity, is that it is assumed to be uncorrelated with within-person variability, and that this feature also constitutes an important point of contact linking the RI-CLPM to the DPM.

**Overview of existing statistical models and related issues**

*LCM-SR*

Besides the stable trait factors, the RI-CLPM and the STARTS model share the feature that they do not emphasize the modeling of trajectory of changes for group or individuals. Indeed, the group mean is expressed as $\mu_{xt}$, $\mu_{yt}$, and the mean structure always perfectly fits to the data (unless we impose assumptions such as equality constraints between time points). Expected scores for individuals given the stable trait factors are expressed as $\mu_{xt} + I_{xi}$, $\mu_{yt} + I_{yi}$. The trajectories of each individual and group are mutually parallel, and these models do not intend to describe trajectories and their individual differences in a structured manner using some common (e.g., growth) factors. In other words, the difference between these two models is therefore characterized not by their mean structures, but by their covariance structures.

By contrast, in descriptive research using longitudinal data, the primary interest lies in understanding the average trajectory of change and individual differences in trajectory. From this perspective, the latent growth model (LCM) is a method that emerged within psychometrics, with factor analysis as one of its intellectual lineages. Moreover, proposals for integrative statistical models began to appear mainly from the 2000s onward. These models, like the LCM, use common factors (referred to as growth factors) to provide a parsimonious representation of the average trajectory and individual differences, while at the same time, like the CLPM, allowing inferences about the reciprocal relations among variables through autoregressive and cross-lagged terms.

The LCM with structured residuals (LCM-SR) (Curran, Howard, Bainter, Lane, & McGinley, 2013; Chi & Reinsel, 1989) is one such statistical model, and was proposed as a method for inferring within-person relations. The LCM-SR has a representation that separates the part describing the trajectory of change from the regression equation including the autoregressive and cross-lagged terms. Specifically, in an LCM-SR assuming a trajectory of linear change, observations $x_{it}$ and $y_{it}$ are expressed like the LCM as

$$\begin{aligned} x_{it} &= I_{xi} + (t-1)S_{xi} + x_{it}^*, \\ y_{it} &= I_{yi} + (t-1)S_{yi} + y_{it}^*, \end{aligned} \quad (6a)$$

where $I_{xi}$ and $I_{yi}$ are intercepts of trajectory for individual *i* or, more specifically, intercept factors reflecting the magnitude of the true value at the first time point ($t = 1$). $S_{xi}$ and $S_{yi}$ are slope factors representing the magnitude of the slope of individual *i*'s trajectory. Unlike the stable trait factor, these two factors are estimated without fixing their means to 0. We also assume and estimate the covariance of these factors. A positive covariance indicates that the larger the true value at the first time point (the intercept), the larger the subsequent changes (the slope). In the context of the LCM, the intercept and slope factors are sometimes collectively referred to as the growth factors.

$x_{it}^*$ and $y_{it}^*$ are the residual terms (or, corresponding to the previous expressions, the deviations), which can also be described as the quantity obtained by detrending the observations, that is, by removing their linearly expressed trend component. As in the (RI-)CLPM, at $t \geq 2$, the residual terms are further expressed using regression equations with lagged variables, such as

$$\begin{aligned} x_{it}^* &= \beta_{xt} x_{i(t-1)}^* + \gamma_{xt} y_{i(t-1)}^* + d_{xit}^*, \\ y_{it}^* &= \beta_{yt} y_{i(t-1)}^* + \gamma_{yt} x_{i(t-1)}^* + d_{yit}^*, \end{aligned} \quad (6b)$$

and $y_{i1}^*$ are treated as exogenous variables, their (co)variances are estimated, and they are further assumed to be uncorrelated with the growth factors. This reflects the intention of the LCM-SR, similar to the RI-CLPM, to capture within-person relations by using residual terms that are separated from and assumed to be uncorrelated with the trajectories. Note that if we assume time-varying autoregressive coefficients, cross-lagged coefficients, and residual (co)variances on the right side of the regression equation, longitudinal data with at least $T = 4$ will be required for model identification.

LCM-SR may seem suitable for inferring within-person relations. However, within-person relations expressed as Eq. (6b) are relations of the residuals ($x_{it}^*$ and $y_{it}^*$) "after" the individual's trajectory of change is controlled by the growth factors. In particular, if one controls for the slope factor $S$, which is often regarded as reflecting within-person change over time and individual differences therein—both of which are likely to be major components for the reciprocal relations in many cases—the within-person relations of substantive interest are likely to be distorted. This indicates over-adjustment known in causal inference literature. This means that while the intent is to control and detrend a quantity (the slope factor score $S$) that is considered unnecessary for inferring within-person relations, we are also unjustifiably excluding quantities that are necessary for the inference (components reflecting within-person change and their individual differences), which Usami, Murayama et al. (2019) describe as "throwing the baby out with the bathwater".

Even if separating the slope factor from Eq. (6b) can be justified, a potential limitation in applying the LCM-SR is that the shape of the trajectory (e.g., linear) must be specified correctly. In other words, if model misspecification occurs and the actual mean structure is not accurately represented, thereby producing biased estimates, and such bias will be transmitted to the bias in cross-lagged coefficients that lie at the core of inferences regarding within-person relations.

For these reasons, it is the author's position that, when the goal is to infer within-person relations, the RI-CLPM should generally be regarded as having an advantage over the LCM-SR. As noted above, however, the actual data generating process and the true model are unknown. If the $(t − 1)S$ term does not reflect the main components of the reciprocal relations of interest, but rather serves as a time-varying confounder summarizing various unobserved confounding factors, then the use of the LCM-SR may be justified (Usami, Murayama et al. 2019). Moreover, it is clear that the within-person relations represented by the RI-CLPM in this case cannot be interpreted to reflect causal relations, because such time-varying confounders are not controlled for. However, it also seems rare that such a simplified function can describe actual time-varying confounders.

*LCS*

One motivation for the proposal of the latent change score (LCS) model, or the latent difference score model, by McArdle and Hamagami (2001) was to provide, like the LCM-SR, an integrated extension of the LCM and the CLPM. Unlike the LCM-SR, this model was not originally proposed with the explicit objective of inferring within-person relations, but it includes autoregressive and cross-lagged terms and is used to examine reciprocal relations between variables.

In the original LCS representation, as the model's name indicates, the amounts of change (of true values) between time points ($\Delta_{xit} = f_{xit} - f_{xi(t-1)}$) is explicitly introduced and regression equation is specified for this change. However, as in Usami, Hayes, and McArdle (2015, 2016) and Usami, Murayama et al. (2019), a mathematically equivalent representation can be given without using this change, so from the perspective of comparison with other statistical models, the following explanation is based on expressions that do not explicitly use amounts of change.

As with the STARTS model, LCS decomposes the observation into a true score $f$ and measurement error ε, as

$$x_{it} = f_{xit} + \varepsilon_{xit}, \qquad y_{it} = f_{yit} + \varepsilon_{yit}. \qquad (7a)$$

Because of the identification reason, time-invariant error (co)variances are usually assumed. Unlike the statistical models described above, a regression equation that includes autoregressive and cross-lagged terms is set directly on the true scores instead of deviations as

$$\begin{aligned} f_{xit} &= A_{xi} + \beta_{xt} f_{xi(t-1)} + \gamma_{xt} f_{yi(t-1)} + d_{xit}, \\ f_{yit} &= A_{yi} + \beta_{yt} f_{yi(t-1)} + \gamma_{yt} f_{xi(t-1)} + d_{yit}. \end{aligned} \qquad (7b)$$

Also, a common factor A is included in this equation. LCS usually assumes time-invariant autoregressive coefficients and cross-lagged coefficients, due to scale invariance with respect to the measurements described below, but here we assume time-varying coefficients for the sake of model comparison. Furthermore, because improper solutions occur relatively frequently in the LCS, strong assumptions are sometimes imposed to avoid them, such as fixing the residual (co)variances to zero (i.e., $d_{xit} = d_{yit} = 0$). When time-varying autoregressive coefficients, cross-lagged coefficients, and residual (co)variances are assumed, longitudinal data with at least $T = 4$ waves are required for model identification.

The common factors $A_{xi}, A_{yi}$ are random intercepts in the regression equations and are related to the amount of change between time points for individual $i$ (e.g., Usami et al., 2015). Although both are referred to as "intercepts", $A_{xi}$ and $A_{yi}$ are not equivalent to the growth factors ($I_{xi}$ and $I_{yi}$) in the LCM-SR, because they are included in the equations together with the autoregressive and cross-lagged terms. By contrast, the means of $A_{xi}$ and $A_{yi}$ are not fixed to zero; rather, their means and (co)variances are estimated, as is the case for growth factors. In addition, the true scores at the initial time point ($f_{xi1}$ and $f_{yi1}$) are treated as exogenous variables, and their means and (co)variances are estimated. Moreover, correlations (covariances) between these true scores and the common factors $A_{xi}$ and $A_{yi}$ are also assumed and estimated. This assumption of correlation stands in contrast to the assumption in the RI-CLPM and STARTS that the stable trait factors and deviations (within-person variability), or in the LCM-SR that the growth factors and deviations (within-person variability), are uncorrelated.

So how does the common factor *A* differ from the stable trait factor *I*? This paper primarily focuses on the RI-CLPM, but the only mathematical difference between the RI-CLPM and the STARTS model is the assumption of measurement errors, so here we will compare the LCS and the STARTS model, both of which include measurement errors. Another feature shared by these models is that, aside from measurement errors, only a single common factor (*I* or *A*) is included in the model. Unlike the stable trait factor *I* (and likewise, the growth factor in the LCM-SR), the common factor *A* is included in the regression equation along with the autoregressive and cross-lagged terms. This difference concerns how each common factor contributes to the true scores (or observations). In the STARTS model, as shown in Eq. (5b), the stable trait factor contributes only as a direct effect on the true score at each time point, for example in the form $I_{xi} \Rightarrow f_{x3}$. Moreover, the magnitude of this effect is given by $I_{xi}$ or $I_{yi}$ and is constant across time points. The growth factors in the LCM-SR also share this property (this is the reason why the same symbol, *I*, is used for both intercept factors and stable trait factors).

By contrast, with regard to the common factor *A* in the LCS, consider, for example, its contribution from $A_x$ to $f_{x3}$. Given the recursive relation in Eq. (7b), it is clear that there exists not only a direct effect such as $A_x \Rightarrow f_{x3}$, but also an indirect effect such as $A_x \Rightarrow f_{x2} \Rightarrow f_{x3}$ through the autoregressive path. Furthermore, unlike the stable trait factor *I*, $A_x$ can also affect a different variable (*Y*) through the cross-lagged path, as in $A_x \Rightarrow f_{x2} \Rightarrow f_{y3}$.

More generally, this means that the effect of the common factor $A_x$ on $f_{xt}$ propagates to the true score at the next time point, $f_{x(t+1)}$, through the autoregressive path, and then further to $f_{x(t+2)}$. In this way, the contribution of $A_x$ to the true score (or observation) at a given time point accumulates over subsequent time points. As a result, unlike a stable trait factor, the magnitude of the contribution of $A_x$ differs across time points. Usami, Murayama et al. (2019) referred to a common factor with this property as an *accumulating factor*, in order to distinguish it from a stable trait factor. Reflecting this difference, the STARTS model and the LCS have different covariance structures and therefore yield different estimates of the cross-lagged coefficients. Moreover, comparison of the path diagrams for the dynamic panel model (DPM; Figure 1d), which, like the LCS, includes an accumulating factor, and the RI-CLPM (Figure 1b) makes the characteristics of the accumulating factor described above visually clearer.

This difference between accumulating factors and stable trait factors implies mathematical and conceptual differences in between-person heterogeneity being controlled for, and Eq. (7b) therefore cannot be regarded as expressing within-person relations equivalent to those represented in the RI-CLPM or the STARTS model. Because the magnitude of the contribution varies over time, what is controlled for by an accumulating factor cannot be considered as a stable trait. Although such time-varying property in themselves is flexible and appealing, it seems difficult to judge whether the effects of (time-invariant) confounders in actual longitudinal data can be adequately represented through a relatively simplified function such as that in Eq. (7b).

More importantly, unless time-invariant constraints are imposed on the parameters, the LCS does not satisfy scale invariance. In other words, if the observed variables are linearly transformed, not only the means of the accumulating factors but also the estimates of the autoregressive and cross-lagged coefficients, as well as the model fit, will change. This implies that the autoregressive and cross-lagged coefficients in the LCS are not separated from the trajectory of change (i.e., the mean structure). By contrast, if time-invariant parameters are assumed in order to ensure scale invariance, the model may become unrealistic, especially when examining within-person relations over a long period of time, because qualitative differences may arise in the dynamics of change.

Statistical models such as the LCM-SR and LCS, which are intended to enable the simultaneous modeling of trajectories of change (i.e., mean structures) and the estimation of reciprocal relations as within-person relations, may appear reasonable in that they seem to achieve two aims at once. However, because these are not, in fact, mutually independent research hypotheses (Usami, Murayama et al., 2019), attempting to investigate them simultaneously may instead hinder the investigation of each hypothesis, with the result that one who chases two hares may catch neither. By introducing growth factors ($S$) or accumulating factors ($A$) into the model instead of the group mean $\mu$, which serves only to capture the mean structure, there is a risk that the within-person relations that one truly wishes to capture through the cross-lagged coefficients will be distorted. Therefore, when one is interested in both modeling trajectories of change and inferring within-person relations, it would be preferable to examine them separately, for example by using the LCM and the RI-CLPM.

Another model proposed with the intention of providing an integrative extension of the LCM and the CLPM is the autoregressive latent trajectory model (ALT; Curran &

Bollen, 2001). Ignoring measurement errors, the ALT corresponds to an LCS model with an additional accumulating factor ($B$) that has time-varying weights (factor loadings) (Usami, Murayama et al., 2019). For the same reasons discussed above, it is highly likely that, in the ALT as well, the cross-lagged coefficients do not reflect the within-person relations that one actually wishes to capture, and they therefore cannot be regarded as representing the within-person relations equivalent to that of the RI-CLPM.

### *GCLM*

The general cross-lagged panel model (GCLM) was proposed in the field of organizational research by Zyphur, Allison et al. (2020) and Zyphur, Voelkle et al. (2020) with the aim of developing a method that would come closer to causal inference through a generalization of the CLPM. Among the statistical models discussed thus far, it is the only one proposed after Hamaker et al. (2015) and Usami, Murayama et al. (2019), and although it has not been applied as extensively as the RI-CLPM, a considerable number of empirical applications have already been reported. Zyphur, Allison et al. (2020) position the GCLM as a statistical model that can represent the dynamics of change more flexibly by including "stable trait factors" [6] as unit effects and moving-average (MA) terms.

Without loss of generality, we explain the GCLM assuming first-order lag. For $t \geq 3$, this model can be represented as[7]

$$x_{it} = \alpha_{xt} + \lambda_{xt} B_{xi} + \beta_{xt} x_{i(t-1)} + \gamma_{xt} y_{i(t-1)} + \delta_{xt} d_{xi(t-1)} + \zeta_{xt} d_{yi(t-1)} + d_{xit},$$
$$y_{it} = \alpha_{yt} + \lambda_{yt} B_{yi} + \beta_{yt} y_{i(t-1)} + \gamma_{yt} x_{i(t-1)} + \delta_{yt} d_{yi(t-1)} + \zeta_{yt} d_{xi(t-1)} + d_{yit}, \quad (8)$$

where $\alpha_{xt}$ and $\alpha_{yt}$ are intercept terms representing the effect at time point $t$ (occasion effect), $\beta_{xt}$ and $\beta_{yt}$ are the autoregressive (AR) coefficients, and $\gamma_{xt}$ and $\gamma_{yt}$ are the cross-lagged (CL) coefficients. $B_{xi}$ and $B_{yi}$ are common factors referred to by Zyphur, Allison et al. (2020) and Zyphur, Voelkle et al. (2020) as unit effects or "stable trait factors," and their factor means are fixed at zero. The initial observations ($x_{i1}$ and $y_{i1}$) are treated as exogenous variables, and their means and (co)variances are estimated. In addition, although not stated explicitly by Zyphur, Allison et al. (2020) and Zyphur, Voelkle et al. (2020), it is assumed—similarly to the stable trait factors in the RI-CLPM—that the common factors $B_{xi}$ and $B_{yi}$ are uncorrelated with the initial observations $x_{i1}$ and $y_{i1}$ (however, because of the nature of $B$ as an accumulating factor, discussed below, they are correlated with the observations from the second time point onward, and this assumption itself can also be relaxed). $\lambda_{xt}$ and $\lambda_{yt}$ are weights (factor loadings). Although Zyphur, Allison et al. (2020) and Zyphur, Voelkle et al. (2020) allow these weights to be fixed at 1, they recommend assuming time-varying weights.

The terms $\delta_{xt}$ and $\delta_{yt}$ are moving-average (MA) coefficients representing the effects of residuals from the same variable at the previous time point, whereas $\zeta_{xt}$ and $\zeta_{yt}$ are cross-lagged moving-average (CLMA) coefficients representing the effects of residuals from a different variable at the previous time point. MA terms have traditionally been widely used, especially in models for time series data with a large number of time points. In the GCLM, effects from the same variable are represented by AR and MA terms, whereas effects from a different variable are represented by CL and CLMA terms. When time-varying coefficients and the residual (co)variances are assumed, longitudinal data with at least $T = 4$ waves are required for model identification.

Zyphur, Allison et al. (2020) and Zyphur, Voelkle et al. (2020) interpret the common factor *B* in the sense of the stable trait factor (*I*) in the RI-CLPM, and explain that the model obtained by removing the MA and CLMA terms from the GCLM is equivalent to the RI-CLPM (that assumes time-varying weights for the stable trait factor). However, as is clear from the discussion thus far in this paper and from the form of Eq. (8), the common factor *B* is in fact an accumulating factor. In this sense, the GCLM does not provide a representation of within-person relations equivalent to that of the RI-CLPM (Usami, 2021). Moreover, although an accumulating factor with time-varying weights may provide a more flexible representation from the standpoint of model fit, it is also likely to overcontrol, as in the LCM-SR, information on within-person change and individual differences therein—major components of reciprocal relations—thereby distorting the within-person relations that one actually wishes to examine.

In addition, Usami (2021) pointed out that the residual terms associated with the MA and CLMA terms are generally vague in terms of what they substantively represent, and that, because they are virtually controlled for multiple times (e.g., through both the AR and MA terms, or through both the CL and CLMA terms), the meaning of the cross-lagged coefficients in the GCLM becomes more complicated and may also induce multicollinearity. Of course, whether it is appropriate to include MA and CLMA terms depends on the true data-generating process, and the complexity of interpretation of coefficients does not in itself imply inferential bias or reduced estimation accuracy; nonetheless, these remain practical problems that the GCLM may entail.

***Summary of this subsection and a unified framework for the models***

The LCM-SR, LCS (or ALT), and GCLM proposed in psychology and related fields all introduce common factors in order to represent and control for latent confounders and, more broadly, between-person heterogeneity. These models are characterized primarily by the type of common factor they include (i.e., a stable trait factor or an accumulating factor), the presence or absence of a mean structure for that factor, and the presence or absence of correlations with within-person variability. Although there is the fundamental difficulty that the true data-generating process is usually unknown, the potential problems in inferences about within-person relations discussed in this subsection can be classified into three types: problems of overadjustment (LCM-SR, GCLM), problems arising from attempting to infer within-person relations and trajectories of change (mean structures) simultaneously (LCM-SR, LCS), and complexity in the interpretation of coefficients (GCLM).

The LCM-SR, like the RI-CLPM, enables inference regarding within-person relations through deviations that are assumed to be uncorrelated with the growth factors ($I$, $S$); however, because it may overadjust for the slope factor $S$, which is usually considered to reflect within-person change and individual differences therein—major components of reciprocal relations—the within-person relations one actually wishes to examine are likely to be distorted, and the cross-lagged coefficients may therefore be biased. In the LCS, the accumulating factor $A$ is included in the regression equations together with the autoregressive and cross-lagged terms, and as a result, unlike a stable trait factor, it contributes to the true scores (or observations) in a time-varying manner. Although this feature is flexible, it is difficult, for example, to judge whether the effects of (time-invariant) confounders in actual longitudinal data can be adequately represented through a simplified function such as that specified by the model. In addition, because the LCS does not include the group mean $\mu$, it fails to satisfy scale invariance unless time-invariant constraints are imposed on parameters such as the autoregressive and cross-lagged coefficients. The GCLM, which includes the accumulating factor $B$, likewise carries a risk of overadjusting for information on within-person change and individual differences, as in the LCM-SR, because it introduces an accumulating factor with time-varying weights (factor loadings). Moreover, the inclusion of moving-average terms whose substantive meaning is obscure may complicate the interpretation of the cross-lagged coefficients and may also induce multicollinearity.

These models are also considered to carry a higher risk of improper solutions than the RI-CLPM (e.g., Orth et al., 2021). Although Orth et al. (2021) did not directly examine the GCLM, the complexity of its structure suggests that it may be at least as prone to improper solutions as the LCM-SR and LCS, if not more so.

Since Hamaker et al. (2015), models other than the RI-CLPM—particularly the GCLM—have increasingly been applied. As discussed above, however, the meaning of the within-person relations represented by these models is not the same. Moreover, because these models also provide mathematically different mean structures and covariance structures, they yield different estimates of the cross-lagged coefficients as well (e.g., Usami, Murayama et al., 2019; Orth et al., 2021; Usami, 2021). In actual applications, however, it seems that insufficient attention has often been paid to these differences and to the potential problems discussed above.

Besides mathematical differences such as the type and number of common factors, including stable trait factors and accumulating factors, differences in the notation and terminology adopted in the literature introducing each model also made it difficult to distinguish clearly among the existing models. Usami, Murayama et al. (2019) presented a unified framework for explaining the mathematical and conceptual differences among various statistical models used to infer reciprocal relations from three perspectives: (i) the type of common factor included (i.e., whether it is an accumulating factor), (ii) whether there is an interest in modeling trajectories of change (mean structures), as indicated by whether terms such as the group mean $\mu$ or intercept $\alpha$ are included, and (iii) whether measurement errors are assumed. They further showed, through path diagrams and conceptual figures, that each model can be positioned as a special case within this unified framework—that is, as a submodel assuming only certain common factors or parameters—and also illustrated the mathematical relationships among the models. Moreover, if the presence of moving-average terms is also taken into account, the GCLM can likewise be characterized within the same framework (Usami, 2021). However, although those studies referred to the issue in connection with accumulating factors, the dynamic panel model (DPM), which has been used mainly in economics, was not directly included as a model for comparison.

**Overview of the DPM and its mathematical relation to the RI-CLPM**

*Overview of the DPM*

Although the DPM (Chigira et al., 2011; Wooldridge, 2011; Hsiao, 2014; Allison, Williams, & Moral-Benito, 2017) is typically used to analyze unidirectional rather than reciprocal relations, for purposes of comparison its formulation for the two-variable case considered thus far may be written as follows (e.g., Allison et al., 2017; Usami, 2023):

$$x_{it} = \alpha_{xt} + A_{xi} + \beta_{xt} x_{i(t-1)} + \gamma_{xt} y_{i(t-1)} + d_{xit},$$
$$y_{it} = \alpha_{yt} + A_{yi} + \beta_{yt} y_{i(t-1)} + \gamma_{yt} x_{i(t-1)} + d_{yit}. \qquad (9)$$

Here, $\alpha_{xt}$ and $\alpha_{yt}$ are intercept terms representing the effects of time point $t$. $A_{xi}$ and $A_{yi}$ are accumulating factors (though in economics they are typically referred to as unit effects or individual effects), and their means are zero. Further, $\beta_{xt}$ and $\beta_{yt}$ are autoregressive coefficients, $\gamma_{xt}$ and $\gamma_{yt}$ are cross-lagged coefficients, and $d_{xit}$ and $d_{yit}$ are residuals. Correlations (covariances) are assumed and estimated between the initial observations ($x_{i1}$ and $y_{i1}$), as well as between those values and the two accumulating factors[8]. In addition, just as the CLPM can be rewritten from Eq. (1) to

Eq. (2), Eq. (9) can likewise be reformulated using the group mean $\mu$ instead of the intercept $\alpha$:

$$x_{it} = \mu_{xt} + x_{it}^*, \qquad y_{it} = \mu_{yt} + y_{it}^*, \tag{10a}$$

$$x_{it}^* = A_{xi} + \beta_{xt} x_{i(t-1)}^* + \gamma_{xt} y_{i(t-1)}^* + d_{xit}^*,$$

$$y_{it}^* = A_{yi} + \beta_{yt} y_{i(t-1)}^* + \gamma_{yt} x_{i(t-1)}^* + d_{yit}^*. \tag{10b}$$

The estimates of the autoregressive coefficients, cross-lagged coefficients, and accumulating factors, as well as the model fit, are unaffected by this reformulation. From a comparative perspective, the path diagram of the DPM under this formulation is shown in Figure 1d.

It should be noted, however, that the term *dynamic panel model* encompasses a wide variety of models, including those with trend terms or moving-average terms. Moreover, compared with the statistical models that have primarily been used in psychology and related fields and discussed thus far, dynamic panel models in economics are often applied in settings with a large number of time points (e.g., exceeding 100 or even 1,000) and relatively short time intervals (lags), such as in analyses of exchange rates. In addition, economists are relatively often interested in forecasting future values. For these and related reasons, it is common in dynamic panel modeling to assume time-invariant coefficients. The same is true for the discussion in Andersen (2022), referred to below, who also uses the term DPM. In the present paper, however, for purposes of model comparison, Eqs. (9) and (10) assume time-varying coefficients, while also taking into account the time-invariant case. It should therefore be noted that, in this paper, the term DPM is used to refer to a limited class of models that do not include moving-average terms and that allow time-varying coefficients.

In addition, in economics, for the DPM (that assumes time-invariant coefficients), estimation often relies on methods other than maximum likelihood, such as the generalized method of moments (GMM), because the form of the estimator depends on whether the unit effects are treated as random effects or fixed effects, as well as on the assumptions regarding the initial conditions discussed below (Chigira et al., 2011). However, because the primary focus of this section is on comparing the formulations of different statistical models, here I assume the use of maximum likelihood based on SEM (ML-SEM), as with the preceding models. For typical model specifications and notations for dynamic panel models, as well as details of ML-SEM implementation and comparisons of estimation results with GMM, see Moral-Benito (2013) and Allison et al. (2017).

Now, ignoring measurement errors, the DPM in Eq. (9) corresponds to the LCS when the intercept term $\alpha$ is included and the mean of the accumulating factor is fixed to zero. As a result, because the mean structure can be expressed as a function of $\alpha$, the problem of scale invariance observed in the LCS with time-varying coefficients is resolved. From another perspective, the DPM can also be understood as corresponding to the GCLM when the time-varying weights associated with the accumulating factors are fixed at 1 and the MA and CLMA terms are set to 0. Therefore, although this ultimately depends on the true data-generating process, the risks of overadjustment and problems of interpretability are likely to be substantially reduced.

*Mathematical relation with the RI-CLPM*

For these reasons, the DPM may be regarded as enabling inference regarding within-person relations in a more appropriate way than the models discussed in the previous subsection (LCM-SR, LCS, and GCLM). This then raises the question of which of the two models—the DPM or the RI-CLPM, which has rapidly gained popularity in psychology and related fields—should be considered preferable as a statistical model.

To address this question, it is first necessary to clarify the mathematical relationship between these models. At first glance, they may appear mathematically unrelated, because the RI-CLPM in Eq. (3) is formulated in terms of deviations, whereas the DPM in Eq. (9) is formulated in terms of observations. However, it is known that, provided the autoregressive and cross-lagged coefficients are time-invariant (and the absolute values of the eigenvalues of the coefficient matrix are less than 1; e.g., Hamaker, 2005), the two models yield mathematically equivalent representations and the same coefficients when certain assumptions are added to each model (Andersen, 2022). One such assumption is that, in the DPM, to take into account that the process of change has already been operating prior to $t = 1$, a constraint is imposed on the weights (factor loadings) associated with the accumulating factors at this time point by a function of the parameters; under this constraint, the DPM becomes equivalent to the RI-CLPM. This type of specification is sometimes referred to as the constrained approach (e.g., Jongerling & Hamaker, 2011). It has also long been known that, when the coefficients are time-invariant, imposing analogous constraints on the weights in the ALT by the same procedure renders it equivalent to the LCM-SR (Hamaker, 2005; Usami, Murayama et al., 2019).

The other assumption is that, if correlations between the stable trait factors and the initial observations are allowed in the RI-CLPM, it becomes equivalent to the DPM (Andersen, 2022). The estimates of the (time-invariant) cross-lagged coefficients also become identical. This type of specification is sometimes referred to as the predetermined approach. Following Andersen (2022), the present paper refers to the RI-CLPM under this specification as the predetermined RI-CLPM, and its path diagram is shown in Figure 1c. As noted in the previous section, under this specification the stable trait factors, like accumulating factors, become common factors that do not exert time-invariant direct effects, and the variance of the observations can no longer be orthogonally decomposed into within-person and between-person components. Therefore, although introducing such correlations gives rise to the equivalence between the predetermined RI-CLPM and the DPM, the predetermined RI-CLPM should no longer be regarded as the same kind of statistical model as the RI-CLPM, but rather as one based on a different design logic. Moreover, when time-varying coefficients are assumed, the equivalence between the predetermined RI-CLPM and the DPM no longer holds. In that case, in addition to the RI-CLPM, there exist at least two candidate statistical models (i.e., predetermined RI-CLPM and the DPM) that involve accumulating (or, "accumulating-like") factors.

Table 1 presents the results of applying the predetermined RI-CLPM and the DPM to the MACC data explained earlier. As in the previously reported analyses of the CLPM and RI-CLPM, estimation was carried out using ML-SEM. The table shows that, when time-invariant autoregressive coefficients, cross-lagged coefficients, and residual (co)variances are assumed, the predetermined RI-CLPM and the DPM yield identical estimates of the coefficients (as well as their standard errors). By contrast, this relationship does not hold when time-varying coefficients are assumed. In addition, reflecting the fact that these two models offer more flexible representations in the sense that their accumulating (or accumulation-like) factors contribute in a time-varying manner, they show better fit than the RI-CLPM. Moreover, under the time-varying condition, the DPM exhibits somewhat different estimation results from those of the RI-CLPM and the predetermined RI-CLPM, in that the statistical significance of some of the cross-lagged coefficient estimates is not consistent across the models.

**Reconsidering the RI-CLPM: The role of the stable trait factor from the view of causal inference and model selection**

In light of the foregoing discussion, I reconsider what the stable trait factor controlled for in the RI-CLPM is, what it means from the perspective of causal inference, and how model selection should be conducted.

**Stable trait factor, confounder, and centering**

In the RI-CLPM, the stable trait factor contributes to the observed variables only as a (time-invariant) direct effect, while it does not contribute to the deviations, that is, the within-person variability. In the terminology of causal inference, the observed variable is a collider that is the joint consequence of both the stable trait factor and the within-person variability, and the path from the stable trait factor to within-person variability is blocked by the observed variable, which functions as a collider.

From this perspective, Usami (2023), while explicitly presenting the data-generating process that reflects the RI-CLPM (Figure 1b) and the identification conditions for causal effects, argued that although the stable trait factor controls for unobserved between-person heterogeneity, it cannot be regarded as a (time-invariant) confounder like an accumulating factor[9]. However, the within-person relations represented in the RI-CLPM are relations among deviations, which are latent variables, and inference regarding them must ultimately be based on information from the observed variables. Therefore, ignoring and failing to account for the stable trait factor that links the observed variables and the within-person variability information, despite its presence, induces bias in the estimation of the cross-lagged coefficients.

This suggests that, if time-invariant confounders are in fact present in the data-generating process, the RI-CLPM cannot directly accommodate them, and their influence is only partially absorbed into the stable trait factors. The inclusion of the stable trait factors should therefore be understood not so much as controlling for a confounder, but rather as corresponding to the operation of centering by the (true) person mean, as is often discussed in the context of hierarchical linear modeling (e.g., Usami, 2017; Asparouhov & Muthén, 2018). Moreover, centering may be viewed as the key operation that links models for deviations such as the RI-CLPM (or LCM-SR) and models for observations such as the DPM (or ALT). Although Wang and Maxwell (2015) discussed the decomposition of within-person and between-person relations and centering in hierarchical linear models (that do not include lagged variables), further investigation is needed regarding the relation between the interpretation of the common factors in the various statistical models discussed in this paper and whether centering is or is not applied to each variable.

**Data-generating process and assumptions about initial conditions**

Usami (2023) also explained that the treatment of the variable at the initial time point ($t = 1$), referred to as the initial condition, differs depending on whether the common factor assumed in the data-generating process is a stable trait factor or an accumulating factor. In survey and observational studies in particular, it is natural to assume that, in many longitudinal datasets, the actual data-generating process has already been operating prior to $t = 1$. In such a case, if an accumulating factor is assumed in the process, its influence has already accumulated in the observations at the initial time point, and therefore modeling that takes this influence into account is required, as in the LCS, ALT, or DPM (for example, by using the predetermined approach, which assumes correlations between the initial observations and the accumulating factor; e.g., Gische, West, & Voelkle, 2021). By contrast, if only a stable trait factor is assumed in the process, it contributes only as a direct effect on the observations and does not exert a cumulative influence on within-person variability at the initial time point. Therefore, even if the process had already been operating before $t = 1$, there is no need to assume a correlation between the stable trait factor and the deviation (or within-person variability) at $t = 1$ (Usami, 2023). In other words, if it is plausible to assume that the process does not contain heterogeneity in the form of a confounder like an accumulating factor, then the inclusion of a stable trait factor uncorrelated with within-person variability, together with the orthogonal decomposition into within-person variance and between-person variance as in the RI-CLPM, may be regarded as a valid procedure.

**Critiques of the RI-CLPM and model selection**

Lüdtke and Robitzsch (2022), through simulation and analytical investigations, argued that when unobserved time-invariant confounders are present, the conditions under which the RI-CLPM can properly deal with their influence are quite limited. As discussed above, this reflects the fact that the stable trait factor controlled for in the RI-CLPM is not a confounder and contributes only in the form of a time-invariant direct effect. Lüdtke and Robitzsch (2022) also argued the usefulness of the CLPM with higher-order lags. However, whether this is appropriate from the perspective of causal inference again depends on the true data-generating process, and as the CLPM contains neither common factors nor unit effects, the range of (unobserved) heterogeneity that it can directly capture is likely to be limited.

Furthermore, Lüdtke and Robitzsch (2022) criticized the inference of within-person relations based on variability around the within-person mean (expected value), arguing that such inference is established only by ignoring the potentially important influences of factors explaining between-person differences and therefore departs from the investigation of the original causal hypothesis. Relatedly, Orth et al. (2021) recommended the use of the CLPM for inferring between-person relations and the RI-CLPM for inferring within-person relations. As Lüdtke and Robitzsch (2022) themselves noted, these issues can ultimately be understood as concerning differences in the estimand, that is, the target of estimation corresponding to the scientific question of interest, which has recently been emphasized in biostatistics and epidemiology. As Hamaker et al. (2015) originally pointed out, the relations examined in the CLPM cannot usually be clearly characterized as either a pure within-person relations or a between-person relations. Therefore, when one is interested in inferring within-person relations, it is generally preferable to introduce common factors (or unit effects) under some assumptions, as in the RI-CLPM or DPM, so as to separate inference concerning within-person relations from that concerning between-person relations.

In causal inference, it is generally required that the specified statistical model appropriately reflects the assumed data-generating process (e.g., Gische et al., 2021). However, in most cases it is unknown whether the true data-generating process contains a stable trait factor, an accumulating factor, both, or neither, and through what paths or functional forms these contribute to the observations (e.g., Curran & Bauer, 2011; Andersen, 2022; Usami, 2023). Moreover, although an accumulating factor is flexible in that it can represent time-varying contributions to observations that are not limited to direct effects, it is not necessarily easy to judge whether the influence of (time-invariant) confounders in actual longitudinal data can be appropriately represented through the paths or functional forms assumed, for example, in the DPM.

Taken together, when one is interested in inferring within-person relations, one realistic approach may be to estimate a particular statistical model—such as the RI-CLPM, the DPM, or the predetermined RI-CLPM—based on the assumed data-generating process, while also estimating other candidate models as sensitivity analyses, as in the comparison shown in Table 1, and examining the appropriateness of model selection and the robustness of the estimation results with reference to fit indices, information criteria, and residual correlations (e.g., Usami, 2023).

## Summary and future directions

### Summary of this paper

Seven years have passed since Hamaker et al. (2015) criticized the CLPM, and the RI-CLPM has rapidly spread in psychology and related fields with such momentum that it may replace the CLPM, long regarded as the gold standard. Although this paper has presented several issues surrounding the RI-CLPM, at least in the comparison between the CLPM and RI-CLPM, the CLPM—which does not include a stable trait factor reflecting unobserved between-person heterogeneity—cannot be regarded as suitable for inferring within-person relations, even if higher-order lags are assumed. By contrast, although the RI-CLPM appears to represent between-person heterogeneity in a somewhat limited way, it can nevertheless be positioned as one effective method for inferring within-person relations.

At the same time, however, many other statistical models are also available. Moreover, although the precise meaning of the within-person relations represented by each model and the estimates of the cross-lagged coefficients usually differ across models, the mathematical and conceptual relations among them have not necessarily been clearly organized, and in actual applications insufficient attention seems to have been paid to these differences. Although the discussion of statistical models for inferring reciprocal relations as within-person relations has become increasingly complex and diverse, with many contributions continuing to appear, in this paper I have attempted to organize these issues, drawing also on my own work (Usami, Murayama et al., 2019; Usami, 2021, 2023).

Because the actual data-generating process and the true model are generally unknown to the researcher, it is difficult to judge whether the common factors or unit effects included in a statistical model appropriately control for latent confounders (more broadly, between-person heterogeneity). This makes it difficult to arrive at a definitive conclusion regarding the best model choice for inferring within-person relations and the proper interpretation of the cross-lagged coefficients. Against this background, this paper has aimed to clarify the potential problems that may arise in inferring within-person relations from existing statistical models (LCM-SR, LCS, ALT, and GCLM) from the perspectives of overadjustment, the problem of simultaneously inferring within-person relations and trajectories of change (mean structures), and the complexity in interpretation of coefficients.

A statistical model that avoids these problems and includes an accumulating factor capable of representing time-varying contributions to observations that are not limited to direct effects is the DPM expressed in Eqs. (9) or (10). By contrast, the stable trait factor in the RI-CLPM contributes to the observations only as a time-invariant direct effect representing stable individual differences. The RI-CLPM and the DPM may both be strong candidate statistical models for inferring within-person relations, but it cannot necessarily be stated clearly which of them more accurately captures the actual data-generating process and yields less biased inference regarding within-person relations. Although the accumulating factor in the DPM is flexible, it is not necessarily easy to judge whether the influence of (time-invariant) confounders in actual longitudinal data can be appropriately represented through the paths and functional forms assumed by the model. In addition, this paper has described another model specification, termed the predetermined RI-CLPM, in which correlations are assumed between the initial observations and the stable trait factors, although under this specification the stable trait factors can no longer be interpreted in the same way as those in the RI-CLPM. When time-invariant autoregressive coefficients, cross-lagged coefficients, and residual (co)variances are assumed, the predetermined RI-CLPM and the DPM are mathematically equivalent (Andersen, 2022), but when such assumptions are not imposed they yield different estimation results (Table 1).

Although the available options for statistical models are thus diverse, one realistic strategy may be to estimate a particular statistical model—such as the RI-CLPM, the DPM, or even the predetermined RI-CLPM—based on the assumed data-generating process, while also estimating other candidate models as sensitivity analyses and examining the appropriateness of model selection and the robustness of the estimation results by referring to fit indices, information criteria, and residual correlations (e.g., Usami, 2023).

**Future directions**

RI-CLPM or DPM, or some other statistical model such as the predetermined RI-

CLPM? Regardless of these differences in model choice, however, the need remains unchanged to identify and collect confounders in advance—particularly time-varying confounders—and to control for them appropriately within the statistical model. Many of the statistical models introduced in this paper, including the RI-CLPM, are typically estimated through SEM, and when there are observed confounders they are often represented under the assumption of linear relations with the variables of interest ($X$, $Y$) in the inference of within-person relations. Although such procedures are useful to some extent, especially when continuous variables are analyzed, the linearity assumed in the standard SEM and path analysis has often been criticized from the perspective of causal inference (e.g., Hong, 2015).

At the same time, the causal inference literature has proposed some approaches that do not assume such linearity, as well as robust approaches that are less affected by model misspecification. From this perspective, Usami (2023), under the data-generating process assumed by the RI-CLPM, provided mathematical definitions of the stable trait factor and within-person variability and proposed a method for inferring within-person relations based on stepwise estimation that does not necessarily assume linearity with time-varying confounders. More specifically, scores representing within-person variability (within-person variability scores) are first predicted for each variable through a measurement model based on SEM (or factor analysis), and these are then treated as observed variables. Methods known mainly in epidemiology as marginal structural models (MSMs) and structural nested models (SNMs) (e.g., Robins & Hernán, 2009; Hernán & Robins, 2021) are then applied to estimate the effects. This approach can also be extended to inference under the assumption that an accumulating factor such as that included in the DPM is present. Methodological development for inferring within-person relations, inspired by approaches originating in causal inference and not necessarily confined to psychometrics, is likely to be an important direction for future research.

The problem of separating or distinguishing within-group relations from between-group relations, or within-person relations from between-person relations, has long been of interest in psychology and related fields. In that sense, discussion surrounding reciprocal relations as within-person relations—an old yet new issue—entered a new phase after Hamaker et al. (2015). At present, this topic is being actively discussed, especially in connection with the RI-CLPM, but the discussion is still centered largely among psychometricians and cannot yet be said to involve, for example, researchers in other related fields such as econometrics, epidemiology, or causal inference. I hope that this paper will help organize the various issues surrounding the inference of within-person relations and stimulate new questions, thereby promoting healthier discussion and methodological development among not only psychometricians but also substantive researchers in psychology and related fields who make use of these methods.

# Footnotes

[1] Within-person and between-person relations are sometimes referred to as intra-individual and inter-individual relations, respectively.

[2] However, intensive longitudinal data—characterized not only by a large number of individuals but also by a large number of time points (with the number and timing of measurements often differing across individuals), as in cases where automated assessments from wearable devices are utilized—have also increasingly been collected in recent years. For such data, approaches based on continuous-time models (Voelkle, Oud, Davidov, & Schmidt, 2012; Ryan, Kuiper, & Hamaker, 2018), which treat time in a continuous form, have recently attracted growing attention, because they can address the problem that the magnitude of the (cross-lagged) coefficients may vary depending on time intervals (lags) between time points. The development and comparison of continuous-time models corresponding to the RI-CLPM and the other statistical models discussed in this paper appear to be important topics for future research.

[3] Orth et al. (2024) attempted to establish empirical guidelines for the effect sizes of cross-lagged coefficients obtained when applying the CLPM and RI-CLPM.

[4] The STARTS model originated as an extension of the latent state-trait model, which was designed to decompose a single longitudinally measured variable into trait and state components. Although the STARTS model was initially formulated for a single variable, it was subsequently extended to accommodate two or more variables, at which point cross-lagged terms were incorporated into the model.

[5] Usami, Murayama et al. (2019) used the term "stationarity" to refer to the setting of time-invariant autoregressive coefficients, cross-lagged coefficients, and residual (co)variances.

[6] Zyphur, Allison et al. (2020) referred to stable trait factors in their model as "stable factors".

[7] Zyphur, Allison et al. (2020) and Zyphur, Voelkle et al. (2020) assumed time-invariant autoregressive coefficients $\beta_x$, cross-lagged coefficients $\gamma_x$, moving-average coefficient $\delta_x$, and cross-lagged moving-average coefficient $\zeta_x$, but an extension to time-varying coefficients is also possible.

[8] If accumulating factors are viewed as random intercepts, they may appear similar to the hierarchical linear models commonly used in psychological research for longitudinal data, namely those that include lagged variables and assume random effects. However, hierarchical linear models do not include intercept terms representing time-specific effects, nor do they usually allow coefficients to vary across time points. In addition, time-varying residual variances are not typically assumed, and a major difference is that the random intercept is assumed to be uncorrelated with the explanatory variables. That said, the distinction between random-effects and fixed-effects models is also related to the issue of centering the variables. For a discussion of this point in the case of models without lagged variables, see Hamaker and Muthén (2020).

[9] In Usami, Murayama et al. (2019), the stable trait factor was described as a confounder, but this point was corrected in Usami (2023).

Table 1: Comparison of results across statistical models.

| | CLPM | | | | RI-CLPM | | | | Predetermined RI-CLPM | | | | DPM | | | |
|---|---|---|---|---|---|---|---|---|---|---|---|---|---|---|---|---|
| | time-invariant | | time-varying | | time-invariant | | time-varying* | | time-invariant | | time-varying | | time-invariant | | time-varying | |
| | Est. | SE | Est. | SE | Est. | SE | Est. | SE | Est. | SE | Est. | SE | Est. | SE | Est. | SE |
| $\beta_{y2}$ | **.82** | .01 | **.91** | .03 | **.70** | .01 | .22 | .43 | *.53* | *.02* | **.59** | .05 | *.53* | *.02* | **.45** | .04 |
| $\gamma_{y2}$ | **.04** | .01 | .05 | .03 | -.01 | .02 | .08 | .06 | *-.01* | *.02* | -.04 | .05 | *-.01* | *.02* | -.02 | .03 |
| $\beta_{x2}$ | **.51** | .01 | **.52** | .01 | **.17** | .01 | **.29** | .14 | *.17* | *.01* | **.24** | .02 | *.17* | *.01* | **.20** | .02 |
| $\gamma_{x2}$ | **.01** | .00 | **.05** | .02 | -.01 | .01 | 2.4 | 1.7 | *-.01* | *.01* | -.03 | .02 | *-.01* | *.01* | **.06** | .03 |
| $\beta_{y3}$ | **.82** | .01 | **.85** | .02 | **.70** | .01 | **.63** | .03 | *.53* | *.02* | **.31** | .07 | *.53* | *.02* | **.47** | .03 |
| $\gamma_{y3}$ | **.04** | .01 | **.09** | .03 | -.01 | .02 | .07 | .05 | *-.01* | *.02* | .01 | .05 | *-.01* | *.02* | .03 | .03 |
| $\beta_{x3}$ | **.51** | .01 | **.47** | .01 | **.17** | .01 | **.12** | .02 | *.17* | *.01* | **.14** | .02 | *.17* | *.01* | **.15** | .02 |
| $\gamma_{x3}$ | **.01** | .00 | **.05** | .01 | -.01 | .01 | **.06** | .02 | *-.01* | *.01* | .01 | .03 | *-.01* | *.01* | **.04** | .02 |
| $\beta_{y4}$ | **.82** | .01 | **.85** | .02 | **.70** | .01 | **.70** | .03 | *.53* | *.02* | .11 | .11 | *.53* | *.02* | **.50** | .03 |
| $\gamma_{y4}$ | **.04** | .01 | .04 | .03 | -.01 | .02 | .02 | .05 | *-.01* | *.02* | .14 | .07 | *-.01* | *.02* | .04 | .04 |
| $\beta_{x4}$ | **.51** | .01 | **.53** | .01 | **.17** | .01 | **.11** | .02 | *.17* | *.01* | **.11** | .02 | *.17* | *.01* | **.17** | .02 |
| $\gamma_{x4}$ | **.01** | .00 | -.01 | .01 | -.01 | .01 | .00 | .01 | *-.01* | *.01* | .00 | .04 | *-.01* | *.01* | .00 | .01 |
| $\beta_{y5}$ | **.82** | .01 | **.82** | .01 | **.70** | .01 | **.76** | .02 | *.53* | *.02* | **.58** | .04 | *.53* | *.02* | **.54** | .02 |
| $\gamma_{y5}$ | **.04** | .01 | -.03 | .03 | -.01 | .02 | -.06 | .05 | *-.01* | *.02* | -.02 | .06 | *-.01* | *.02* | -.02 | .04 |
| $\beta_{x5}$ | **.51** | .01 | **.49** | .01 | **.17** | .01 | **.11** | .02 | *.17* | *.01* | **.08** | .03 | *.17* | *.01* | **.11** | .02 |
| $\gamma_{x5}$ | **.01** | .00 | .01 | .01 | -.01 | .01 | -.01 | .01 | *-.01* | *.01* | .01 | .02 | *-.01* | *.01* | .01 | .01 |
| $\beta_{y6}$ | **.82** | .01 | **.79** | .01 | **.70** | .01 | **.66** | .02 | *.53* | *.02* | **.51** | .03 | *.53* | *.02* | **.53** | .02 |
| $\gamma_{y6}$ | **.04** | .01 | .02 | .03 | -.01 | .02 | -.06 | .05 | *-.01* | *.02* | -.01 | .05 | *-.01* | *.02* | .03 | .04 |
| $\beta_{x6}$ | **.51** | .01 | **.55** | .02 | **.17** | .01 | **.20** | .03 | *.17* | *.01* | **.18** | .03 | *.17* | *.01* | **.17** | .02 |
| $\gamma_{x6}$ | **.01** | .00 | .00 | .01 | -.01 | .01 | -.01 | .01 | *-.01* | *.01* | .01 | .01 | *-.01* | *.01* | .01 | .01 |
| $Cov(I_y, I_x)|Cov(A_y, A_x)$ | | | | | **.05** | .01 | **.04** | .01 | **.05** | .01 | **.04** | .01 | **.03** | .01 | .01 | .01 |
| $Cov(y_1, x_1)$ | **.05** | .02 | **.04** | .02 | -.02 | .01 | -.01 | .01 | **-.05** | .02 | **-.04** | .02 | .03 | .02 | **.04** | .02 |
| $Cov(I_x, y_1)|Cov(A_x, y_1)$ | | | | | | | | | -.02 | .01 | -.01 | .02 | **.04** | .01 | .01 | .01 |
| $Cov(I_x, x_1)|Cov(A_x, x_1)$ | | | | | | | | | .01 | .01 | -.01 | .01 | **.13** | .01 | **.12** | .01 |
| $Cov(I_y, y_1)|Cov(A_y, y_1)$ | | | | | | | | | **-.60** | .05 | **-.55** | .06 | **.26** | .02 | **.30** | .02 |
| $Cov(I_x, x_1)|Cov(A_x, x_1)$ | | | | | | | | | **.04** | .01 | **.04** | .02 | **.04** | .01 | **.04** | .01 |
| $V(I_x)|V(A_x)$ | | | | | **.15** | .01 | **.15** | .01 | **.15** | .01 | **.15** | .01 | **.10** | .01 | **.40** | .01 |
| $V(I_y)|V(A_y)$ | | | | | **.58** | .04 | **.61** | .03 | **1.1** | .04 | **1.1** | .06 | **.25** | .03 | **.63** | .03 |
| $V(x_1)$ | **.41** | .01 | **.41** | .01 | **.25** | .01 | **.26** | .01 | **.25** | .01 | **.27** | .01 | **.41** | .01 | **.10** | .01 |
| $V(y_1)$ | **.57** | .03 | **.62** | .03 | .03 | .03 | .01 | .01 | **.64** | .06 | **.59** | .06 | **.58** | .03 | **.26** | .03 |
| df | 68 | | 40 | | 65 | | 49 | | 61 | | 33 | | 61 | | 33 | |
| CFI | 0.875 | | 0.895 | | 0.959 | | 0.966 | | 0.964 | | 0.991 | | 0.964 | | 0.988 | |
| AIC | 73465.64 | | 73192.02 | | 72224.17 | | 72125.29 | | 72151.05 | | 71778.16 | | 72151.05 | | 71818.56 | |
| BIC | 73607.51 | | 73514.47 | | 72385.39 | | 72389.69 | | 72338.07 | | 72145.75 | | 72338.07 | | 72186.15 | |
| RMSEA | 0.077 | | 0.091 | | 0.045 | | 0.047 | | 0.043 | | 0.030 | | 0.043 | | 0.034 | |
| SRMR | 0.113 | | 0.101 | | 0.066 | | 0.060 | | 0.066 | | 0.030 | | 0.066 | | 0.031 | |

*Under the "time-invariant" condition, equality constraints are imposed across time points on the autoregressive coefficients, cross-lagged coefficients, and residual (co)variances. However, under the time-varying condition for the RI-CLPM, equality constraints are imposed on the residual (co)variances because of improper solutions. The estimates of the residual (co)variances and the means/intercepts are omitted. Boldface indicates statistical significance at the two-tailed 5% level, and italics indicate that the corresponding values are identical across models.

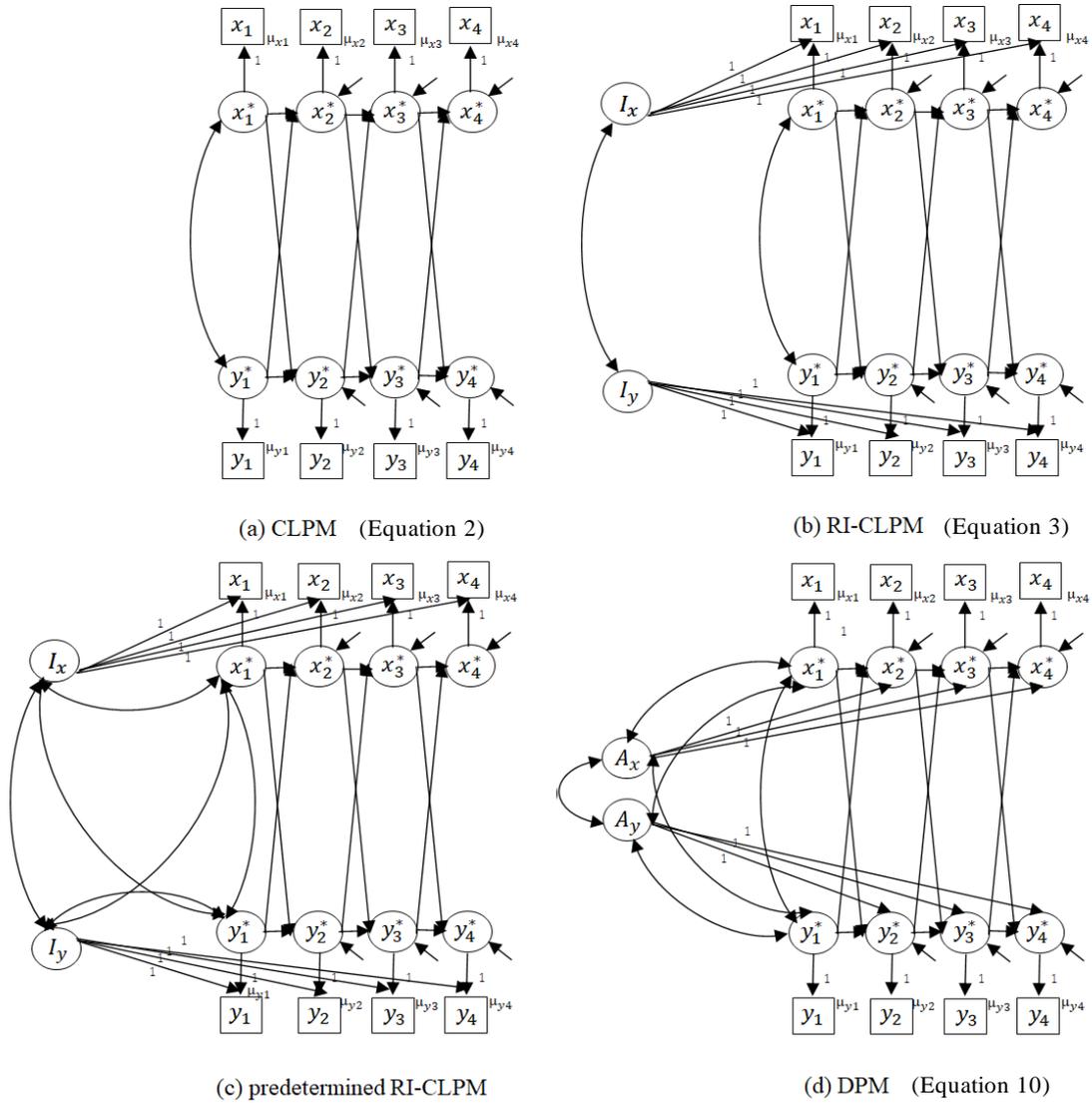

Figure 1: Path diagrams of the statistical models.

*Paths representing residual covariances are omitted. The mean of each common factor is fixed at zero, and its (co)variance is estimated. In (c), the common factor ($I$) in the predetermined RI-CLPM differs in both conceptual and mathematical role from the stable trait factor in the RI-CLPM shown in (b), which exerts time-invariant direct effects on the observations; nor does this common factor orthogonally decompose the variance of the observations into between-person and within-person variances. However, when time-invariant autoregressive coefficients, cross-lagged coefficients, and residual (co)variances are assumed, the corresponding estimates in (c) the predetermined RI-CLPM and (d) the DPM are identical.